\documentclass[twoside,a4paper,11pt]{article}
\usepackage{irrj2e}

\usepackage{lastpage}
\usepackage{hyperref}
\usepackage{xurl}
\usepackage[utf8]{inputenc}
\usepackage{booktabs}
\usepackage{amsmath}
\usepackage{amsfonts}
\usepackage{amssymb}
\usepackage[inline]{enumitem}
\usepackage{enumitem}
\usepackage{mathbbol}
\usepackage[T1]{fontenc}
\usepackage{mathtools}
\usepackage{multirow}
\usepackage{graphicx}

\newcommand{\ie}{\emph{i.e.}}
\newcommand{\eg}{\emph{e.g.}}

\newcommand{\etc}{\emph{etc.}}
\newcommand{\vs}{\emph{vs. }}

\irrjheading{1}{2025}{47--\pageref{LastPage}}{10/24; Revised 03/25}{03/25}{10.54195/irrj.19654}
\ShortHeadings{Search and Society: Reimagining Information Access for Radical Futures}{Mitra}
\firstpageno{47}
\title{Search and Society:\\\Large Reimagining Information Access for Radical Futures}
\author{\name Bhaskar Mitra
  \email bmitra@microsoft.com \\
  \addr Microsoft Research \\
  Montréal, Canada}
\editor{Monica Lestari Paramita}

\begin{document}
\maketitle

\begin{abstract}
Information retrieval (IR) research must understand and contend with the social implications of the technology it produces.
Instead of adopting a reactionary strategy of trying to mitigate potential social harms from emerging technologies, the community should aim to proactively set the research agenda for the kinds of systems we \emph{should} build inspired by diverse explicitly stated sociotechnical imaginaries.
The sociotechnical imaginaries that underpin the design and development of information access technologies needs to be explicitly articulated, and we need to develop theories of change in context of these diverse perspectives.
Our guiding future imaginaries must be informed by other academic fields, such as human-computer interaction,
information sciences, media studies, design, science and technology studies, social sciences, humanities, democratic theory, and
critical theory, as well as legal and policy experts, civil rights and social justice activists, and artists, among others.
In this perspective paper, we motivate why the community must consider this radical shift in how we do research and what we work on, and sketch a path forward towards this transformation.
\end{abstract}

\section{Introduction}
\label{sec:intro}
Information retrieval (IR) research must understand and contend with the social implications of the technology it produces.
Nearly half a century ago, \citet{belkin1976some} concluded that IR research should acknowledge its responsibility to society and ``must become both theoretically self-conscious and self-consciously based upon a social ideology''.
This perspective has gained traction in the IR community in recent years.
Researchers in attendance at the third Strategic Workshop in Information Retrieval in Lorne (SWIRL)~\citep{culpepper2018research} identified fairness, accountability, confidentiality, and transparency in IR (``FACT IR'') as socially consequential and strategically important research directions for the field.
The following year, the FACTS-IR workshop~\citep{olteanu2021facts} added ``safety'' as a fifth pillar.
Subsequently, a large body of recent IR literature has grappled with questions of fairness, transparency, and explainability in the context of information access.

However, it is our perspective that this growing focus on fairness and ethics in IR---despite having played a critical role in bringing much-needed attention to the societal implications of IR systems and advancing the conversation about the IR community's responsibility to broader society---operates within a severely constrained frame that leaves the many underlying values, politics, and socioeconomic incentives that guide IR research largely unchallenged.
For example, faced with the applications of generative artificial intelligence (AI) for information access the IR community has focused on concerns of fair ranking and representation and limiting model ``hallucinations''\footnote{
We acknowledge that the term ``hallucination'' anthropomorphizes AI models and its usage should be discouraged.
However, given the popular usage of that term in the IR community, we make an exception here for clarity.}
for good reasons, but have largely ignored other significant consequences of these technologies on society, such as for the information ecosystem and how these systems concentrate power and control~\citep{mitra2024sociotechnical}.
In machine learning (ML), there has been similar recent perspectives~\citep{blodgett2020language, miceli2022studying} that, for example, calls for shifting the lens from fairness and bias to the power differentials that exists between those who build technology, those who use it, and those who are subject to it.
Others in the ML community have brought attention to the questions of how these technologies shift power~\citep{kalluri2020don} and simultaneously constrain ethics interventions in practice~\citep{widder2023s} and shape our collective sociotechnical futures.
Even by accepting the frame that we should develop fairness and transparency mechanisms for certain systems, we may inadvertently ignore the alternative perspective that some of these technologies should be dismantled, not made fairer, nor more transparent~\citep{barocas2020not, wilkinson2023theories, merchant2023blood}.
Consequently, at the recent fourth SWIRL workshop\footnote{\url{https://sites.google.com/view/swirl2025}}, researchers in attendance called for expanding the ``FACTS-IR'' framing to center IR research on societal, democratic, and emancipatory values.
Similar sentiments, including re-centering IR on societal needs and informing IR research with democratic theories, were also discussed at the first Search Futures workshop~\citep{azzopardi2024search}.

\paragraph{What do we propose?}
In this paper, we argue that IR research, instead of adopting a reactionary strategy of trying to mitigate potential social harms from emerging technologies by developing new fairness and transparency interventions, should aim to proactively set the research agenda for the kinds of systems we \emph{should} build inspired by diverse explicitly stated sociotechnical imaginaries.
Towards that goal, IR research needs to explicitly articulate the sociotechnical imaginaries~\citep{jasanoff2009containing, jasanoff2015dreamscapes} that underpin the design and development of information access technologies, and develop theories of change~\citep{weiss1995nothing, brest2010power, taplin2012theory, wiki:TheoryOfChange}.
\citet{jasanoff2015dreamscapes} define \emph{sociotechnical imaginaries} as ``collectively held, institutionally stabilized, and publicly performed visions of desirable futures, animated by shared understandings of forms of social life and social order attainable through, and supportive of, advances in science and technology''.
These shared visions do not only imagine but also co-produce our futures through development and government of digital technologies~\citep{mager2021future}.
Diverse imaginaries promoted by different corporations, professional communities, political organizations, and social movements can coexist ``in tension or in a productive dialectical relationship''~\citep{jasanoff2015dreamscapes}.

In technological research and development, these diverse, and often diverging, perspectives and visions that guide the community frequently remain implicit and unstated~\citep{wilkinson2023theories} despite the significant influence they exert on what the community focuses on and produces.
Because of the consequential role that access to information plays in political participation by citizens in democratic societies and social transformation~\citep{higgins2013information, polizzi2020information, goldstein2020informed, correia2002information, gonzalez2021better}, and as a social determinant of health~\citep{moretti2012access} and economic progress~\citep{yu2002bridging, mutula2008digital}, it is even more important to critically reflect on the values and motivations that guide the design and deployment of popular information access systems.
\emph{What sociotechnical futures do IR researchers and system designers envision and how do they influence the design of current and future IR systems?
Whose sociotechnical imaginaries are granted normative status and what myriad of radically alternative futures are we overlooking?}
For example, what are the implications of the reliance of popular search and social media platforms on advertising as the primary source of revenue generation~\citep{ang2022how} and how does Big Tech's~\citep{oremus2017big} increasing dominance over academic research~\citep{whittaker2021steep} influences and/or homogenizes the kinds of IR systems we build?
\emph{What is our role, as IR researchers, to safeguard communities from falling victim to crisis of imaginations~\citep{haiven2014crises} and how do we become more open to welcoming influences from radically new sociotechnical imaginaries?}
For example, what would IR systems look like if designed for futures informed by feminist, queer, decolonial, anti-racist, anti-casteist, anti-ableist, and abolitionist thoughts, and if the focus of IR research was not to prop up colonial cisheteropatriarchal capitalist structures but to dismantle them?
We believe that explicating, critiquing, and consciously choosing the values and sociotechnical imaginaries that shape IR research is critical to realizing positive social outcomes through IR research.
As \citet{benjamin2024imagination} argues, exercising our imagination is ``an invitation to rid our mental and social structures from the tyranny of dominant imaginaries'', or as Le Guin put it:

\begin{quote}
  ``The exercise of imagination is dangerous to those who profit from the way things are because it has the power to show that the way things are is not permanent, not universal, not necessary.''
  \begin{flushright}
  -- Ursula K. Le Guin\\
  \emph{The Wave in the Mind: Talks and Essays on\\the Writer, the Reader, and the Imagination}~\citep{le2004wave}
  \end{flushright}
\end{quote}

For IR to concretely support diverse sociotechnical imaginaries the research community also needs to develop their own theories of change.
Theory of change~\citep{weiss1995nothing, brest2010power, taplin2012theory, wiki:TheoryOfChange} can be defined as a participatory process whereby groups and stakeholders articulate their long-term goals and identify necessary preconditions in a planning process.
Consequently, in IR adjacent fields---\eg, in the Fairness, Accountability, and Transparency (FAccT)\footnote{https://facctconference.org} community---there has been similar recent calls to reimagine our sociotechnical futures~\citep{dahora2024better} and develop theories of change~\citep{wilkinson2023theories} to make explicit the visions for desired futures of responsible computing and the strategic pathways that lead to those desired futures.
Here, we argue that IR research similarly needs to both explicitly articulate and support diverse imagined futures and develop corresponding theories of change for how new information access technologies can take us towards these desired worlds.
Theories of change in this context may benefit IR research by encouraging community members to make their goals and assumptions explicit, making it feasible to test stated theories, encouraging the community to work towards building consensus, and even aid in developing potential means of evaluation of desired progress~\citep{weiss1995nothing}.

Maximizing social good and minimizing harm in this context should not just be the concerns of the few in our community working on fairness, transparency, ethics, and related areas, but the domain of all IR research that should be guided by theories of change towards these envisioned futures.
In this context, we largely agree with the perspective of \citet{belkin1976some} but diverge on a critical point which is that we believe IR research should not
seek any singular notions of ``universal'' social ideology but explicitly adopt pluralistic humanistic and emancipatory values and make space for diverse visions and perspectives.

Consequently, the task for IR researchers here is not to put themselves in positions to pick the guiding social ideologies nor push technosolutionism to address today's social problems.
Instead, our guiding future imaginaries must be co-developed with scholars from diverse fields such as human-computer interaction (HCI), information sciences, media studies, design, science and technology studies (STS), social sciences, and political sciences, as well as legal and policy experts, civil rights and social justice activists, and artist, to name a few.
Not all sociotechnical imaginaries are equal in this respect, and what futures guide our research must be informed by the values and ethics of our community, which should be constantly discussed, debated, and challenged by the community as part of our sociotechnical research and be open to external critique.
To summarize, as a research community we should invest our energies and resources to:
\begin{enumerate*}[label=(\roman*)]
    \item Nurture digital spaces where radical visions and projects for human emancipation, social progress, and equity and justice can take shape,
    \item encourage experimentation within our research community with new approaches to information access informed by new sociotechnical imaginaries cross-pollinating through interdisciplinary scholarship, and
    \item ensure that the tools and artefacts we produce as a community do not uphold systems of oppression nor contribute towards other systemic social harms.
\end{enumerate*}

\medskip\noindent
So far, we have argued that IR research should explicate the sociotechnical futures we want to realize and develop theories of change towards these desired futures.
In Section~\ref{sec:background} we present a brief overview of existing literature on fairness and ethics in IR, and share our critical perspectives on it to motivate our work.
Then, in Section~\ref{sec:proposal} we draw from relevant movements in IR-adjacent fields, specifically those with explicit values and prefigurative politics; we share some ideas on how the community can get started on this journey; and who should be doing this work.
In Section~\ref{sec:why} we motivate why now is an appropriate time for the community to consider this shift.
We conclude in Section~\ref{sec:conclusion} with some final remarks on potential pitfalls and desired outcomes.
Our goal with this paper is to raise sociopolitical consciousness in the IR research community so that we \emph{all} see our research embedded in projects of future world making and recognize our collective responsibility to affect social good; and to dismantle the artificial separation between the work on fairness and ethics in IR and the rest of IR research.
\section{Background}
\label{sec:background}
IR systems act as intermediaries between information seekers and information artefacts.
These artefacts may represent:
\begin{enumerate*}[label=(\roman*)]
    \item economic and other opportunities for consumers,
    \item monetization opportunities for content creators and publishers,
    \item specific sociopolitical frames and ideologies, and
    \item lenses to view individuals and groups that are subjects of representation by the content.
\end{enumerate*}
These systems infer the information need from highly incomplete expressions of interests (\eg, short keyword-based search queries for web search) or implicit signals (\eg, history of previously accessed artefacts in case of recommender systems), and make subjective estimates of an artefact's relevance to the information need.
Consequently, these systems are not neutral tools for lookup~\citep{noble2018algorithms} and the choices these systems make exert systemic influence over what information is exposed and consumed at scale.
These systems bear a responsibility to society to not only mitigate potential harms, like allocative and representational harms~\citep{crawford2017trouble}, but also to maximize social good.

Representational harms may happen due to reinforcement of negative stereotypes (\eg, by disproportionately suggesting arrest record searches in ads corresponding to searches for black-identifying first names~\citep{sweeney2013discrimination} or suggesting racist stereotypes in query autocompletion~\citep{noble2018algorithms}), by pandering to the white male gaze (\eg, by sexualizing women of color in search results~\citep{noble2018algorithms, urman2024foreign}), and through erasure (\eg, underrepresenting women and other historically marginalized peoples in image search for occupational roles~\citep{kay2015unequal}).
Allocative harms may manifest from disparate exposure in search and recommendation results~\citep{singh2018fairness}---\eg, when women are recommended lower-paying jobs in ads~\citep{datta2014automated} or by influencing traffic to websites that depend on ad-monetization.
Beyond direct representational and allocative harms, these systems also hold tremendous power to shape political discourse and culture~\citep{grimmelmann2008google, gillespie2019algorithmically, hallinan2016recommended}.

In light of these, there has been multiple calls~\citep{culpepper2018research, olteanu2021facts} in the IR community to study and address these potential harms.
\citet{bernard2023systematic} report a significant rise in publications in this area after 2016, with fairness and transparency receiving the most attention.
The increasing focus on these sociotechnical aspects of information access has been at least partly in response to recent advances in foundation models~\citep{bommasani2021opportunities} and their implications for the future of information access~\citep{shah2022situating, mitra2024sociotechnical}.

Fairness in ranking has garnered so much interests that there are numerous recent surveys~\citep{ekstrand2021fairness, zehlike2021fairness, zehlike2022fairness1, zehlike2022fairness2, pitoura2022fairness, dinnissen2022fairness, aalam2022evaluation, patro2022fair, wang2023survey, li2023fairness, deldjoo2023fairness} and tutorials~\citep{ekstrand2019fairness, gao2020counteracting, li2021tutorial, fang2022fairness, bigdeli2022gender} summarizing this emerging body of work, as well as shared tasks~\citep{biega2020overview, biega2021overview, ekstrand2023overview}.
The fairness questions have typically been framed around disparate quality-of-service---\eg,~\citep{mehrotra2017auditing, mehrotra2018towards, neophytou2022revisiting, wu2024towards}---and disparate exposure---\eg,~\citep{asia:equity-of-attention, singh2018fairness, singh:fair-pg-rank, diaz2020evaluating, zehlike2020reducing, PatroBGGC20FairRec, wu2022joint}.
Several recent works~\citep{smith2022recsys, raj2020comparing, raj2022measuring, boratto2022consumer, boratto2023consumer} have also systemically compared various fairness metrics proposed in the literature.

Beyond fairness, there has been renewed interests in questions of transparency and explainability~\citep{zhang2020explainable, anand2022explainable}, addressing misinformation~\citep{zhou2018fake, kumar2018false, sharma2019combating, collins2021trends, zhou2020survey, saracco2021overview, guo2022survey}, and broader ethical concerns in IR~\citep{schedl2022retrieval}.
Transparency in IR covers a broad range of scenarios and concerns.
Examples include transparency about how the system behaves~\citep{singh2019exs, verma2019lirme, singh2018posthoc, zhuang2020interpretable} and how data subjects are represented in search results~\citep{biega2017learning, li2022exposing}.
Transparency needs may specifically arise in the context of how information and knowledge access systems modulate \emph{what} and \emph{who} get exposure and influence how we see ourselves and others~\citep{cortinas2024through}.
Different notions of transparency may be relevant here, including but not limited to: System transparency (\ie, how does the system work?), procedural transparency (\ie, in what social norms and processes is the system use embedded?), or transparency of outcomes (\ie, what is the impact of the system's use on individuals and society?).
New transparency needs~\citep{liao2023ai} may also arise in the context of emerging technologies, such as large language models (LLMs)~\citep{openai2023gpt4, thoppilan2022lamda, chowdhery2022palm, touvron2023llama}.

\subsection{Critical perspectives on fairness and ethics research in IR}

It is important that we critically assess how the body of fairness and ethics research in IR translates to real world impact.
But on this, there is little in the published literature to go on.
It is plausible, for example, that the research on ranking fairness has been operationalized in, or at least has influenced the designs of, popular IR systems in recent years, but institutions who build these systems have rarely publicly disclosed any information in that regard, may be to obfuscate details of system design from bad actors or for competitive reasons.
Alternatively, it is also plausible that in fact many of these approaches have \emph{not} been operationalized, or only been operationalized in very constrained settings, because industry adoption is lagging behind research or that existing fairness research is built on abstractions and assumptions that are incompatible with real world deployment.

Studies of search logs~\citep{jiang2013mining, chuklin2015click} have historically served an important role in IR research for understanding user and system behaviors and how the two interact.
Similarly, online experimentation~\citep{kohavi2007practical, kohavi2009online, kohavi2020trustworthy} has been key to validating the outputs of IR research in the real world. 
In contrast, to the best of our knowledge, there are very few fairness studies---\eg,~\citep{mehrotra2017auditing, mehrotra2018towards, raj2023patterns}---in IR that make use of search logs, while it may be argued that the user interaction data in these logs are exactly where we \emph{should} be looking to identify which social harms are common in practice and understand how exactly they manifest.
Similarly, there is an urgent need for validating proposed fairness interventions from the literature through online experimentation, involving real users and real information needs, to ensure that fairness research is grounded in actual needs of the society and does not amount to just academic intellectual exercises.
While the lack of log-based studies and online experimentation in fairness research is likely due to the lack of access to commercially-deployed systems and corresponding log data for academic research, we must critically enquire why this manifests so much more severely in fairness research compared to other areas of IR, such as in ranking.
To do so, we must expand the frame beyond questions of algorithmic fairness, and examine the very sociopolitical context in which this research is being conducted, the economic incentives and risks that shape it, and the power differentials between institutions and individuals that determine what research is allowed and who is allowed to do it~\citep{whittaker2021steep, widder2023s}.
This lack of access to data and systems not only makes it difficult to reproduce, validate, and challenge the claims in existing fairness studies,\footnote{
For example, some studies on log-based fairness audits depend on the availability of user provided demographic attributes.
In our experience, such attributes are often available only for a small subset of users---such as for signed-in users---who typically are more loyal users of the product and on average much better satisfied with system performance than the general population.
Audits based on this sub-population may significantly under-report bias and unfairness issues faced by users.
Better understanding of such practical challenges can motivate the community to work on specific research questions, such as fairness considerations under distributional shifts and with noisy~\citep{ghazimatin2022measuring, mehrotra2022fair} or missing demographic attributes~\citep{lazovich2022measuring, do2022optimizing}.}
but also limits what fairness questions the community is allowed to investigate.

We should also critically reflect on what questions should or should not be framed as fairness problems, and the societal consequences of doing so.
For example, one of the motivating scenarios described by \citet{morik2020controlling} is exposure fairness for search results across different ends of the political spectrum.
A similar question has also been studied by \citet{kulshrestha2017quantifying}.
Casting this as a fairness issue, however, has several problematic implications and consequences.
Firstly, this assumes an overly-simplistic frame in which complicated intersecting political ideologies are mapped to a linear spectrum (\eg, left \vs right and Democrats \vs Republicans) and holds the two ends static rather than a continuously shifting window of acceptable discourse~\citep{lehman2014brief, giridharadas2019america}.
Furthermore, it also assumes that two ends of a political discourse have equal merit and deserve equal exposure, which amounts to \emph{algorithmic bothsidism}.
Finally, in arguing that builders of IR systems should shift exposure out of fairness concerns, it inadvertently normalizes the idea that it is acceptable for institutions and individuals who own these systems to exert enormous influence over public discourse of social and moral import.
Instead, it is our perspective that IR needs a fundamentally different and cross-disciplinary approach to these questions, one that centers on engaging and co-producing with other academic sub-fields, such as STS~\citep{hackett2008handbook} and critical theory of technology~\citep{feenberg1991critical, feenberg2002transforming}.

We must also assess the validity of constructs that we employ in fairness research.
For example, \citet{jacobs2021measurement} point out that race and gender, that are often the focus of group fairness research, are contested constructs, and indeed so is the construct of fairness itself.
While several papers on fairness have employed gender as a group variable, \citet{pinney2023much} caution us that much more care should be taken in this practice, for example, to ensure that we respect everyone's right to self-identify their gender and recognize the fairness concerns of non-binary peoples.
\citet{patro2022fair} encourage us to to move beyond fairness definitions that are grounded in discrete moments and to consider the long-term impact of fairness interventions.
Fairness research itself may contribute towards certain negative externalities in the long-term, such as encouraging more pervasive collection of protected demographic attributes and further intensification of data surveillance~\citep{zuboff2023age} of already marginalized groups in a misguided attempt to bridge the data gaps that may be responsible for the system's disparate quality of service across groups.

Similarly, an important question that all transparency research must contend with is: \emph{transparency towards what end?}
Some works~\citep{polley2021towards, schmitt2022designing} motivate transparency as a means to increase user trust in the system.
However, we should be critical of whether that trust is warranted, or whether transparency mechanisms could in fact draw users into a false sense of safety and distract them from noticing how the system surveils them and subtly manipulates their behavior~\citep{ravenscraft2020spot, morrison2021dark}.
Indeed, \citet{hollanek2023ai} argue that ``only the
sort of transparency that arises from critique---a method of
theoretical examination that, by revealing pre-existing power
structures, aims to challenge them---can help us produce
technological systems that are less deceptive and more just''.
It is that kind of critical reflection that we believe should inform transparency research in IR and lead the community towards explicit goals to challenge power (\eg, the power that system owners hold over users) and safeguard user agency.

On the same lines, misinformation research should be motivated by sociotechnical visions for the future of democratic societies, public health, and knowledge production.
When we broaden that frame, it becomes apparent that the community must not only focus on automated fact checking, an important research problem, but also understand the social, political, and economic conditions under which misinformation and disinformation is produced and disseminated.
The focus of misinformation research then should include identifying, understanding, and addressing the structural mechanisms of misinformation---\eg, data voids~\citep{golebiewski2019data}---as well as ground itself in the articulation of IR's role in online knowledge production, public health education, and information literacy.

Our perspective here on ongoing research on fairness, transparency, and ethics in IR should \textbf{not} be misconstrued as an argument for doing less of this kind of work.
Instead, we believe that the community should be explicit and more ambitious about the changes it wants to affect in broader society and conduct research with a clear mapping between the research goals and the desired social impact.
Just as \citet{johnson2014open} challenges the notion that open data directly leads to information justice, we want the IR community to be cautious in their assumption that working on narrowly defined questions of fairness, transparency, and ethics necessarily contributes towards practical social good.
To be effective in that endeavour, we believe that we should be explicit in articulating our collective visions for our sociotechnical futures, the changes we want to affect in society, and how we envision our research can bring about those changes.

\section{Towards sociotechnical change}
\label{sec:proposal}
Sociotechnical imaginaries are not born in vacuum.
They are moulded and shaped by our values and our politics.
Deliberation over what futures we want to bring into being \emph{is} essentially political, and challenges us to critically reflect on our community's shared, and yet pluralistic, political values.
While the call for explicit political reflection in IR may come as a surprise to some, we need to recognize that our research and the artefacts we produce do not exist outside of the current sociopolitical order~\citep{friedman2007human, flanagan2008embodying, miller2021technology} but as essential cogs in the system, and the absence of political reflection does not imply an absence of politics in our work, but rather translates to implicit complicity in propping up the status quo and neoliberalism~\citep{dourish2010hci, feltwell2018grand, keyes2019human}.
Instead, we can learn from how some of our neighboring fields, \eg, HCI and AI, have engaged with these questions, and reflect on how politics shapes and intersects with our own research agendas.

\subsection{Prefigurative politics in other IR-adjacent fields}
\label{sec:proposal-related}
There are several strands of research in IR-adjacent fields that explicate prefigurative politics~\citep{asad2019prefigurative} and ground research in humanistic~\citep{bardzell2016humanistic, bardzell2015humanistic, werthner2024introduction}, anti-oppressive and emancipatory~\citep{smyth2014anti, bardzell2016humanistic, kane2021avoiding, monroe2021emancipatory, saxena2023artificial}, feminist~\citep{wajcman2004technofeminism, wajcman2010feminist, bardzell2010feminist, bardzell2011towards, bardzell2011feminism, bardzell2016feminist, bardzell2018utopias, d2020data}, queer~\citep{light2011hci, klipphahn2024introduction, guyan2022queer}, postcolonial and decolonial~\citep{irani2010postcolonial, philip2012postcolonial, dourish2012ubicomp, sun2013critical, ali2014towards, ali2016brief, akama2016speculative, irani2016stories, adams2021can, mohamed2020decolonial}, anti-racist~\citep{abebe2022anti}, anti-casteist~\citep{kalyanakrishnan2018opportunities, sambasivan2021re, vaghela2022interrupting, vaghela2022caste, shubham2022caste, kanjilal2023digital}, anti-ableist~\citep{williams2021articulations, sum2024challenging}, anti-fascist~\citep{mcquillan2022anti}, abolitionist~\citep{benjamin2019race, barabas2020beyond, earl2021towards, jones2021we, williams2023no}, post-capitalistic~\citep{feltwell2018grand, browne2022future}, and  anarchist~\citep{keyes2019human, linehan2014never, asad2017creating} epistemologies.
Reviewing this full body of literature is out-of-scope of this work but we briefly present a sample to draw from and motivate new IR research agendas for sociotechnical change.

\citet{bardzell2016humanistic} define humanistic HCI as ``any HCI research or practice that deploys humanistic epistemologies (e.g., theories and conceptual systems) and methodologies (e.g., critical analysis of designs, processes, and implementations; historical genealogies; conceptual analysis; emancipatory criticism) in service of HCI processes, theories, methods, agenda-setting, and practices'', and include emanicipatory HCI as an aspiration of humanistic HCI.
\citet{kane2021avoiding} propose to incorporate emancipatory pedagogy~\citep{freire2020pedagogy} that does ``not advocate the oppressed simply rise and overthrow their oppressors. Instead, [\dots] the oppressors and oppressed create new educational processes that would allow them to work together to create a new type of society that was emancipatory for all''.
%

In STS, there is a body of work~\citep{longino1987can, wajcman1991feminism, hubbard2001science, turkle2004computational, herring2006gender, von2007conceptualising, haraway2013cyborg, michelfelder2017designing} on gendered inequities caused by technology, and how technology and gender relations mutually shape each other~\citep{wajcman2004technofeminism, wajcman2010feminist}.
In HCI, Bardzell et al.~\citep{bardzell2010feminist, bardzell2011towards, bardzell2011feminism, bardzell2016feminist, bardzell2018utopias} propose to incorporate feminist theories~\citep{kolmar1999feminist, friedan2010feminine} into research and practice.
\citet{bardzell2010feminist} posits that feminist theories can contribute to interaction design both by critiquing and by generating new insights that inform and shape designs and design processes.
We can see feminist HCI in practice in the works of Dimond et al.~\citep{dimond2012feminist, dimond2013hollaback}.
In the context of speculative design~\citep{auger2013speculative}, \citet{martins2014privilege} emphasize the need for intersectional~\citep{crenshaw1989demarginalizing, mccall2005complexity, crenshaw2013mapping} feminist lens in critiquing and dismantling structures of oppression.
Feminist theory, methods, and epistemologies have also influenced AI research~\citep{adam1995feminist, adam2013feminist, wellner2020feminist, browne2023feminist, toupin2024shaping}.
\citet{erscoi2023pygmalion} highlight how women are erased from and by AI technologies.
\citet{leavy2021ethical} propose ethical data curation approaches grounded in feminist principles.
Using feminist epistemology, \citet{huang2022ameliorating} critique existing practices of explainable AI, and \citet{varon2021artificial} critique practices for obtaining digital consent in data extractivist practices in AI.
Gender theory have also been employed in these fields in the forms of Queer HCI~\citep{light2011hci} and Queer AI~\citep{klipphahn2024introduction}.
Both highlight \emph{queering}~\citep{star1997politics, brooks2021queering} as a tactic to challenge the basis on which categories are constructed.

\citet{irani2010postcolonial} define postcolonial computing as one that is ``centered on the questions of power, authority, legitimacy, participation, and intelligibility in the contexts of cultural encounter, particularly in the context of contemporary globalization. [\dots] It asserts a series of questions and concerns inspired by the conditions of postcoloniality''.
\citet{avle2017methods} criticize the ``colonizing impulse'' to valorize ``universal methods'' that are supposedly appropriate across cultural and geopolitical boundaries; instead we can draw from works~\citep{alsheikh2011whose, wong2012dao, winschiers2013toward, shaw2014mobile, fox2014community, ahmed2015residual, akama2016speculative, sambasivan2021re} that center on and incorporate indigenous and non-western values and ethics in the critique and development of technologies.
\citet{ali2014towards} argue for decolonial computing over postcolonial which he criticizes as ``Eurocentric critique of Eurocentrism'' that ``tends to privilege cultural issues over political-economic concerns'' and ``is noticeably silent about past injustices and does not
engage with the matter of reparations''.

\citet{kaba2021we} define abolition as ``a long-term project and a practice around creating the conditions that would allow for the dismantling of prisons, policing, and surveillance and the creation of new institutions that actually work to keep us safe and are not fundamentally oppressive''.
The movement challenges us to move beyond the default assumptions and world views of the carceral state and to dismantle the prison-industrial complex.
Incorporating abolitionist values in computing requires us to oppose carceral technologies, surveillance technologies, and military applications~\citep{earl2021towards}.

Post-capitalist computing assumes ``a socio-economic model that completely replaces capital as the primary method of organising society''~\citep{feltwell2018grand}.
Among other subjects, research in this area contends with, ``the racialized dynamics of labor competition''~\citep{irani2018design}, dismantling Big Tech's concentration of power~\citep{verdegem2022dismantling, srnicek2017platform}, and imagining post-work futures~\citep{browne2022future, butler2018interview, srnicek2015inventing}.
Perhaps, the challenges in this area are best summed up in the words of \citet{jameson2003future}: ``it is easier to imagine the end of the world than to imagine the end of capitalism''.

These different political lens lend to imagining new futures of computing but there are some themes that cut across them.
Firstly, they all recognize technology and society as mutually shaping, and reject both technological determinism~\citep{greene2019better} and the frame in which technology exists in, what \citet{pfaffenberger1988fetishised} calls, a fetishised form~\citep{marx1867fetishism} where technology is disembodied and disconnected from social relations.
Secondly, they recognize that the perspectives, goals, and approaches across this spectrum while sometime distinct are also intersecting.
Finally, they all call for structural changes and progress towards alternative futures for society and computing.
Perhaps, these aspirations are best articulated by \citet{keyes2019human}: ``radically reorienting the field towards creating prefigurative counterpower---systems and spaces that exemplify the world we wish to see, as we go about building the revolution in increment''.
To affect said changes we need to both recognize the politics of our work and ground it in broader context of political actions~\citep{wickenden2018reckoning, moore2020towards, green2021data, widder2023s, young2021call}.

\subsection{Proposals for IR}
\label{sec:proposal-reflections}
The survey of works presented in Section~\ref{sec:proposal-related} hopefully provides some seeds of ideas for how IR research can be driven by radical new sociotechnical imaginaries.
This is not to imply that these other IR-adjacent fields have achieved the desired success from these approaches, in fact there are some evidence~\citep{chivukula2020bardzell} that point otherwise.
Rather, we should recognize that how values and politics can inform computing research is still an open question, and they may apply differently to IR than these other fields.
The challenge then for the community is to collectively engage and push towards sociotechnical change.
In the remaining of this section, we discuss how we imagine some of these frames and values can guide us towards open challenges in information access.
However, these examples should be interpreted as just that, as examples, not our recommendation for specific research questions the community should focus on.
The actual research agenda should be developed through participatory processes that simultaneously focuses on both identifying technical research questions and building diverse communities with shared understanding of these challenges and shared commitments to address them.

Through the lens of feminist, queer, and anti-racist IR, we could critique existing approaches to ranking fairness, not only in terms their use of socially constructed categories, such as race and gender~\citep{pinney2023much}, but question if it is the appropriate framing at all for the problems it purports to solve.
For example, instead of trying to algorithmically fix under-representation of women and people of color in image search results for occupational roles, we could reclaim that digital space as a site of resistance and emancipatory pedagogy by allowing feminist, queer, and anti-racist scholars, activists, and artists to create experiences that teach the history of these movements and struggles.

In context of decolonial IR, ongoing fairness research may co-develop relevant local intervention strategies with legal scholars in recognition of significant differences in legal treatment of topics such as \emph{affirmative action} across geographies.
This shifts fairness research away from abstract universal notions of bias and fairness towards locally-significant societal impact (\ie, \emph{think local, act local}).

Anti-oppressive IR research may concern itself with questions such as:
\emph{Can we translate Freire's~\citep{freire2020pedagogy} anti-oppressive pedagogy to strategies for anti-oppressive information access?}
\emph{Can search result pages support dialogical interactions between searchers that allows for communities of searchers to add context to the search results, as an alternative to centralized moderation?}
Unlike conversation search, that is framed as interaction between the user and the system, the idea of dialogical search interfaces challenges us to build sophisticated sociotechnical solutions to support dialog between searchers in context of specific search intents in ways that leads to knowledge production and better digital literacy.
Anti-oppressive and anti-capitalist perspectives may also motivate us to reimagine search and recommender systems as decentralized and federated.

IR research may also employ these lenses as instruments of critique.
For example, in the enterprise context, \citet{gausen2023framework} adopt decolonial and anti-capitalist lens to expose how information and knowledge access systems may commodify and appropriate knowledge from workers.
We should also critically challenge the employment of what \citet{gray2019ghost} calls Ghost Work in IR research both as a labor issue and through the lens of decolonization.
In abolitionist IR, we must ensure that the techonologies we build cannot be used for surveillance or any other military or carceral applications.
The community may also consider more radical direct actions such as developing critical theories of information access, or collectively organizing to abolish Big Tech~\citep{kwet2020digital}.

\subsection{Where do we start?}
\label{sec:proposal-how}
We are calling for not only a significant shift in what the IR community works on, but fundamentally changing the arrangements within our community that determine on an ongoing basis our research agendas.
Beyond explicating our values and sociotechnical imaginaries, we need to develop frameworks that help us appropriately prioritize societal needs against the needs of the user, the publisher, and the platform owners.
We also need new research that reimagines how IR can be informed by different epistemologies and political theories.
Finally, we must also critically reexamine the arrangements within our community and create spaces for shared sense-making in collaboration with those outside of our field.
We elaborate on these further in this section.

In both academic research and industrial deployment, IR places a strong emphasis on the needs of the user (consumer).
This focus motivates various lines of research including: understanding user needs (through lab studies, log analysis, surveys, \etc), optimizing the search system towards those needs (\eg, relevance optimization, personalization, and improving response time), and validating that proposed system changes indeed benefit the user (through online experimentations and further lab studies).
Salient in industry settings are the needs of the system owners---\eg, revenue, market share, and brand---that drive significant decisions for system design and deployment, but have historically been of lesser concern to academic IR research.
Real-world IR system deployments also engage with content producers and publishers, \eg, web publishers and the search engine optimization community for web search engines, and artists for music recommender systems); although how their needs are weighed against the needs of system owners and users may vary, \eg, \citep{guttenberg2018spotify, plaugic2015spotify}.
IR research has considered questions of fairness between producers, but have rarely focused on the power differentials between system owners and producers, and its implications for producers.

\begin{figure}
    \centering
    \includegraphics[width=0.7\columnwidth]{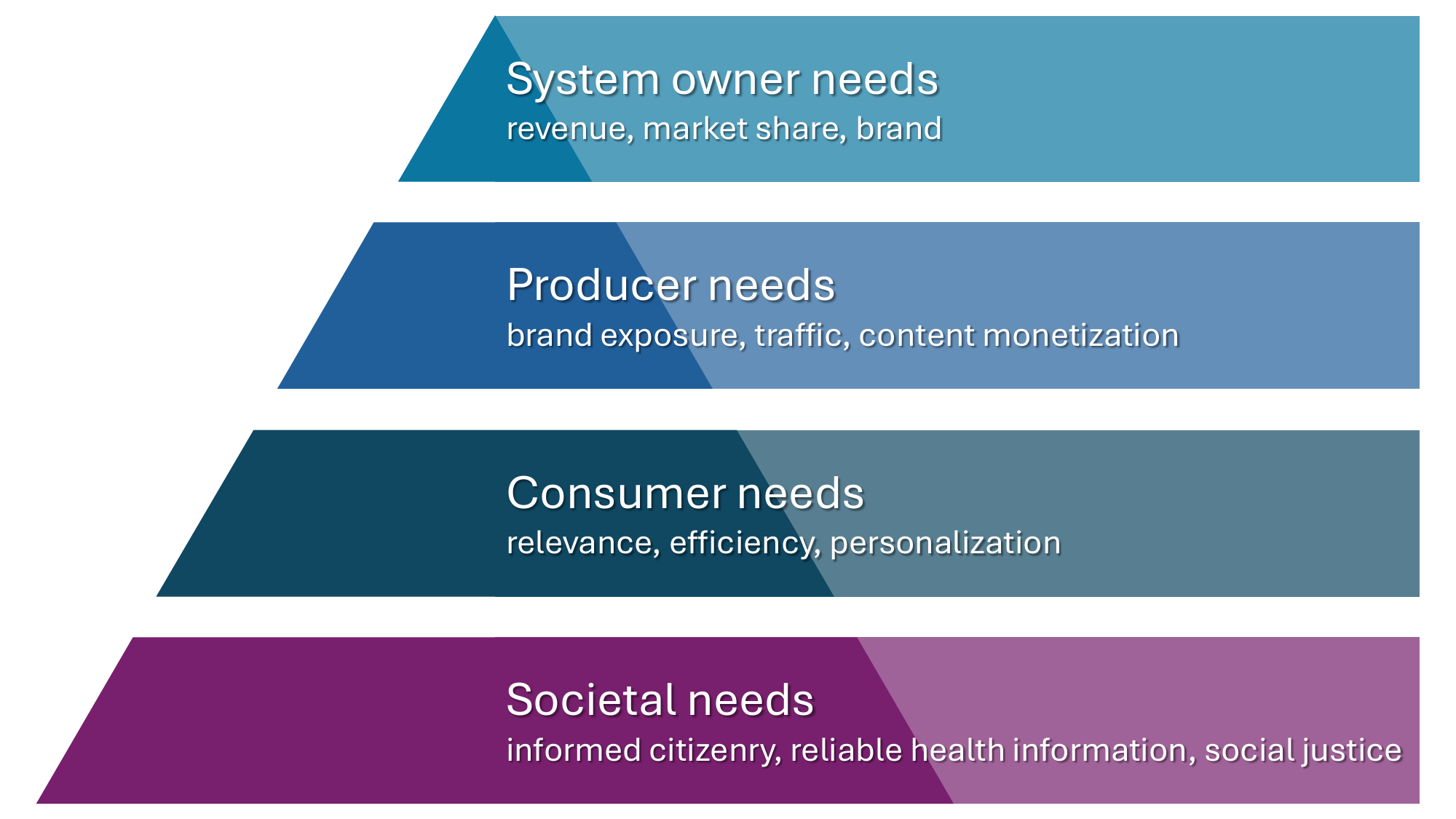}
    \caption{Hierarchy of IR stakeholder needs.
    More fundamental and critical needs are at the bottom of the pyramid.
    This figure is inspired by Maslow's Hierarchy of Needs~\citep{maslow1958dynamic} and Siksika (Blackfoot) way of life~\citep{ravilochan2021blackfoot}.}
    \label{fig:hierarchy-of-stakeholder-needs}
\end{figure}

Societal concerns in both IR research and industry settings have commonly been framed through a narrow lens of harm mitigation, such as ``\emph{how do we make the IR system more fair?}'' and ``\emph{how do we reduce misinformation in search results?}'', without fundamentally challenging the frames in which these systems are designed and deployed, \eg, centralized control and profit incentives~\citep{mager2012algorithmic, taplin2017move}.
IR systems are deeply embedded in sociopolitical and organization context.
However, instead of grounding IR research in questions around its role in online and institutional knowledge production, literacy and informed citizenry, public health education, and social justice, the community has typically constrained themselves to improving measurable system attributes like relevance and efficiency.

Articulating different stakeholder concerns is a prerequisite to any conversation about reprioritizing our research agendas and recentering IR research on societal needs.
In Figure~\ref{fig:hierarchy-of-stakeholder-needs}, we propose a hierarchy of stakeholder needs that IR research should concern with.
Contrary to the status quo, we believe that IR system design and research should explicitly reflect how these systems should contribute to knowledge production, public education, and social movements, and that broader framing of societal concerns should be the most fundamental stakeholder need that should inform and shape IR research.
This should then be followed by concerns of the consumer and producer needs, and lastly the needs of the system owners themselves.
The needs of the consumers, producers, and system owners should not override the need of the collective society, just as the needs of the system owners should not be prioritized over the needs of the consumer and the producers.
This is the shift in our research thinking, agendas, and impact that we are pushing for in this paper.

To realize structural changes in IR and meaningfully challenge dominant imaginaries we must also invest in research that specifically explores and builds on the connections between IR and different epistemologies and social and political theories. 
Examples of such work may consider how IR system design can support different models of democracy~\citep{vrijenhoek2021recommenders, helberger2021democratic} and emancipatory aspirations~\citep{mitra2025emancipatory}.
We must also ensure that such research does not happen in isolation but is grounded in a collective effort to build a movement within the IR community.
This requires us to acknowledge and value community building as an important part of IR research.
In this context, it is important to emphasize that our call for re-centering societal needs in IR research must not be confused for a call for technosolutionism.
Quite the contrary, we believe we need safe spaces where IR researchers can engage with scholars from other diverse fields, legal and policy experts, activists, and artists, in a recognition of a collective struggle to develop shared understandings of the core challenges and what IR research can offer to this process.
These spaces are critical for cross-pollination of ideas and shared sense-making, which are vital for realizing structural changes.
As Tsing aptly puts it:

\begin{quote}
  ``We are contaminated by our encounters; they change who we are as we make way for others. As contamination changes world-making projects, mutual worlds—and new directions—may emerge. Everyone carries a history of contamination; purity is not an option. [\ldots] staying alive—for every species—requires livable collaborations. Collaboration means working across difference, which leads to contamination. Without collaborations, we all die.''
  \begin{flushright}
  -- Anna Lowenhaupt Tsing\\
  \emph{The mushroom at the end of the world:\\On the possibility of life in capitalist ruins}~\citep{tsing2015mushroom}
  \end{flushright}
\end{quote}

Through these exchanges we must build relations of solidarity; work together to articulate pluralistic desirable sociotechnical futures, co-develop theories of change and new research agendas to support our aspirations; vigilantly critique our own assumptions and the structures that we exist in and conduct our research in; and critically assess the impact of our work not by publications or scholarly metrics but in terms of affecting real social change.
In other words, before we can transform our platforms and systems we need to transform our communities and how we conduct research.
And it is vital that we approach these spaces with curiosity and humility; in recognition of our own incomplete understanding of the world; open to change and be changed by these encounters.
In this context, it would do IR researchers good to keep in mind the words of activist Lilla Watson.

\begin{quote}
  ``If you have come here to help me you are wasting your time, but if you have come because your liberation is bound up with mine, then let us work together.''
  \begin{flushright}
  -- Lilla Watson and other members of an\\Aboriginal Rights group in Queensland\footnote{While often credited for this quote, Watson explained~\citep{attributingwords} that this came out of a collective process by an Aboriginal rights group in Queensland that she was part of.}
  \end{flushright}
\end{quote}

This praxis will be unfamiliar and the learning curve frustrating; but hopefully the mutual shaping of IR techonology and society will be ultimately rewarding for us all.
And we must embark on this transformative journey fully acknowledging that challenging dominant imaginaries is challenging power itself.
To affect counter-imaginaries we must therefore actualize counter-structures and alternative funding mechanisms that can sustain this research in the face of likely reprisals from those whose power and visions are threatened by our proposed transformations.

\subsection{Who should do this work?}
\label{sec:proposal-who}
In his highly influential work ``Pedagogy of the oppressed'', \citet{freire2020pedagogy} criticizes the ``banking'' concept of education: ``Education thus becomes an act of depositing, in which the students are the depositories and the teacher is the depositor. [\ldots] knowledge is a gift bestowed by those who consider themselves knowledgeable upon those whom they consider to know nothing. [\ldots] But the humanist, revolutionary educator cannot wait for this possibility to materialize. From the outset, her efforts must coincide with those of the students to engage in critical thinking and the quest for mutual humanization. His efforts must be imbued with a profound trust in people and their creative power. To achieve this, they must be partners of the students in their relations with them.''
These profound words present a critique that we believe is also relevant to how we conduct fairness and ethics research in the IR community today.

Our default modes of doing research, like pedagogy, takes for granted the validity of \emph{experts} and \emph{expertise} as integral to knowledge production.
While this may be effective when our research involves improving ranking or developing new evaluation measures, we posit it is the wrong approach when the goal of the research is to affect sociotechnical change.
It would be narcissistic to imagine that sociotechnical research of moral import can be conducted by the few in our community and then ``deposited'' to the rest.
Instead these concerns should be central to all IR research and we should collectively engage in dialogical praxis.
Ultimately, research that attempts to affect sociotechnical change does not just transform technology, but also the researcher, and both are necessary for progress.

To challenge the homogeneity of the future imaginaries---saliently bound by colonial, cisheteropatriarchal, and capitalist ways of knowing the world---that shape our research, we need broad and diverse participation from our community.
But it is also in that very context that we must critically reflect on the topic of membership in our community itself.
ACM SIGIR has a commendable emphasis on the topic of Diversity, Equity, and Inclusivity (DEI)~\citep{verberne2024report, kobayashi2017opportunities, goharian2022report, goharian2023report, goharian2021women}.
For our sociotechnical imaginaries to be informed by pluralistic social, cultural, and political perspectives we not only need significantly improved representation from historically underrepresented and marginalized groups in our community, but also be inclusive of their politics and world views.
Inclusion of people without inclusion of their history and struggles is simply tokenism and epistemic injustice~\citep{fricker2007epistemic}.
That is why we believe that we, as a community, should go beyond just diversity and inclusivity (D\&I), and enshrine as our goal Justice, Equity, and Diversity \& Inclusivity (JEDI)---in which context D\&I is both a means towards justice and equity, and also an end in itself.
As \citet{keyes2019human} put it: ``This must be about more than just bodies: it is not diversity if we only accept marginalised people who are stripped of the epistemic models that underpin experiences of being Other, or have the work they draw from those models held to an unequal standard of legitimacy''.

Lastly, we reiterate the important role of industry researchers in this process.
They should take advantage of their proximity to large-scale systems to identify, understand, and communicate concerns of societal import and partner with academia to work on those challenges.
The spaces they occupy are also sites for resistance~\citep{wickenden2018reckoning, widder2023s}.

\section{Why now?}
\label{sec:why}
The arguments we present in this paper to reimagine our sociotechnical futures and center IR research on societal needs have always been relevant to the field.
However, there is a confluence of several factors that makes this discourse particularly relevant in the present moment.
Research communities constantly evolve, shaped by ideas and developments from both within the field and adjacent communities, and in response to real world events and changing societal needs.
In case of IR, we believe we are seeing significant developments in both context at present:
\begin{enumerate*}[label=(\roman*)]
    \item A fast-changing landscape across IR and adjacent fields, such as natural language processing (NLP), HCI, ML, and AI, spurred by recent progress in LLMs and other generative AI approaches, and
    \item an increasing recognition of the role of technology, and the communities that build it, in determining our collective futures.
\end{enumerate*}
Consequently, there is a shared sense in the community that right now both IR technologies and IR research have been made malleable and are undergoing transformative changes under these forces of emerging new computational capabilities and evolving societal needs~\citep{azzopardi2024report}.
This presents a timely opportunity for the field to consciously, collectively, and ambitiously engage in purposeful dialog about the future of the field while metaphorically the ``IR(on)'' is hot and before it is irrevocably shaped by unexamined imaginaries of those with power and influence over present day IR research.
In doing so, we must also critically reflect on ``\emph{where do we want to go?}'' (\ie, our sociotechnical imaginaries), ``\emph{how do we get there?}'' (\ie, our theories of change), and ``\emph{who will we go there with?}'' (\ie, our relationships, and that of our work, with other disciplines, governments, industry, and society).
These considerations should drive future IR research as a whole, and we should accept this opportunity to re-center our research agendas on societal needs while dismantling the artificial separation between the work on fairness and ethics in IR and the rest of IR research.

\paragraph{LLMs are changing how we access information.}
The natural language generation capabilities of LLMs are having a profound effect on how we access information and in what context.
Conversational search interfaces have gone from being aspirational~\citep{anand2021dagstuhl, metzler2021rethinking} to being deployed at web-scale (\eg, Bing Chat\footnote{\url{https://chat.bing.com/} (now Microsoft Copilot~\citep{mehdi2023reinventing})} and Google Bard\footnote{\url{https://bard.google.com/}}) in a span of two years.
While the long-term social implications of inserting an LLM between a retrieval system and the information seeker should rightly be met with rigorous skepticism~\citep{shah2022situating}, natural language interfaces are already impacting how we interact with IR systems.
The ubiquitous search box is being challenged as IR becomes more \emph{context-driven} than \emph{user-driven} as a consequence of LLMs increasingly embedding themselves in user's work processes---\eg, Microsoft Copilot for M365~\citep{mehdi2024bringing, warren2024microsoft}---and interacting with the IR system on the user's behalf, under retrieval-augmentation~\citep{lewis2020retrieval, zamani2022retrieval}.

While we should be excited with the new prospects that these emerging AI technologies unlock and recognize that they will shape how we access and interact with information in the future, we must not be duped by AI techno-determinism into believing that there is a single pre-determined path forward.
Instead, we must hold pluralistic views of what IR's future, one that is yet to be determined, looks like and how these technologies will take us there.
In a study of top-cited AI papers, many of which are coauthored by researchers affiliated with industry or elite universities, \citet{birhane2022values} find that the dominant values expressed and operationalized support concentration of power.
So, we must ask: \emph{in what new ways can we imagine accessing and interacting with information, aided by LLMs, if large-scale IR systems were not just a purview of Big Tech?}
\emph{How would LLMs empower us to reimagine IR systems whose explicit goal is to dismantle hierarchies and redistribute power, not to centralize it?}
\emph{What role would AI technologies play in information access that is built explicitly to facilitate dialogical social processes for knowledge production, world building, and our collective struggles for universal emanicipation?}
It is critical that we have these conversations now in the face of ongoing massive technology-driven power shifts in favor of dominant established platforms that grants their visions of the future normative status and shrinks the space for any critique, resitence, or counter-imaginaries.

\paragraph{LLMs are shifting priorities of IR research.}
Over the last decade, deep learning technologies became the new hammer in the toolbox for IR research~\citep{mitra2018introduction, lin2020pretrained, fan2022pre}, dominating IR publications with nearly four out of five papers at the ACM SIGIR 2020 conference being related to deep learning by some estimates~\citep{mitra2020neural}.
One particular focus of neural IR has been on estimating relevance of information artefacts (\eg, documents) to an information intent (\eg, as expressed by a search query) for ranking, a central problem in IR.
Curiously, many of the key ingredients for this research, such as the Transformer architecture~\citep{vaswani2017attention} and the idea of pretrained LLMs, like BERT~\citep{devlin-etal-2019-bert}, came from fields adjacent to IR;
correspondingly, shifting the focus within the field more towards adapting these models for the relevance estimation task---\eg, \citep{DBLP:journals/corr/abs-1901-04085}---and making them more efficient~\citep{frobe2024reneuir}.

More recently, \citet{thomas2023large} demonstrated that LLMs, like GPT-4~\citep{openai2023gpt4}, are able to estimate the actual searcher's preference for documents, given their query, better than several populations of human relevance assessors.
This technology has already been deployed in production at Bing.\footnote{\url{https://twitter.com/IR_oldie/status/1659413086007328768}} 
Putting it bluntly, these LLMs may be getting close to the best we can expect with machine learned general purpose relevance estimators.
If these claims stand the test of time, it may mark a watershed moment for IR research.
Speculatively, we may see the IR community further shifting towards:
\begin{enumerate*}[label=(\roman*)]
    \item Improving efficiency of these models,
    \item focusing on more specialized information needs---\eg, tip-of-the-tongue information needs~\citep{arguello2021tip}, and
    \item increasing investments in measurement and evaluation---\eg, for emerging new IR scenarios, such as retrieval-enhanced machine learning~\citep{zamani2022retrieval}.
\end{enumerate*}
Alternatively, we may ask: \emph{How can the IR community meet this moment, not with apprehension nor with unchallenged exuberance for progress happening in adjacent fields, but truly grasp this opportunity to redefine what it means to work on IR research?}
\emph{Can we be unabashedly discontent with imagining the future of IR based wholly on what new AI progress makes plausible, and instead reimagine our field as a place where knowledge, culture, and radical aspirations meet to demand of technology to make new futures possible?}
Alternatively, if we fail to articulate an aspiring vision for IR research, we risk as a field being reduced to just an application of AI.

\paragraph{LLMs are raising new sociotechnical concerns.}
It is well-known that language models reproduce, and even amplify, harmful stereotypes and biases of moral import~\citep{friedman1996bias} that are present in their training data~\citep{bolukbasi2016man, caliskan2017semantics, gonen2019lipstick, blodgett2020language, bender2021dangers, abid2021persistent}.
One particular mitigation strategy involves using ML approaches that learn from human preferences, such as reinforcement learning from
human feedback (RLHF)~\citep{christiano2017deep, ziegler2019fine}, to align with ``human values''~\citep{kasirzadeh2023conversation, tamkin2021understanding}.
While RLHF has been quite effective in constraining LLMs from producing certain types of offensive and harmful content, we must be wary of any framing of AI ethics, such as \emph{AI alignment}~\citep{russell2015research, gabriel2020artificial, gabriel2021challenge}, that presupposes the existence of \emph{universal values} but that assumption does not hold true in reality~\citep{prabhakaran2022human, birhane2019algorithmic, jobin2019global, png2022tensions, sambasivan2021re}.
This is particularly concerning if we look at this in the context of power asymmetries that exist between powerful private corporations, who have outsized influence over what values these models are optimized for, and those who use these models or are represented in some fashion in the model outputs.
This is further compounded by the lack of appropriate mechanisms for civil society to participate in and challenge these choices.
Indeed, by placing these controls in the hands of the privileged few, we risk further concentration of power.
The concerns of biases in what these models produce, and even biases in what context they refuse to generate~\citep{urman2023silence}, and who gets to influence those decisions have serious implications for information access and society.
Finally, the development of LLMs themselves may involve potential harms to authors~\citep{davis2023sarah, lawler2023ai, browne2024new, shetler2024ai, milmo2024impossible}, crowdworkers~\citep{gray2017humans, ekbia2017heteromation, gray2019ghost, roberts2019behind, jones2020ghost, roberts2021your, williams2022exploited, perrigo2023exclusive, dzieza2023ai}, and even the environment~\citep{bender2021dangers, patterson2021carbon, bommasani2021opportunities, wu2022sustainable, dodge2022measuring, patterson2022carbon}.

In a critical perspective, \citet{shah2022situating} recommend that IR research should focus on developing appropriate guardrails in anticipation of the social implications of these emerging technologies and not be constrained by a singular LLM-powered conversational search vision for IR.
With this we agree, we should be excited by the new capabilities unlocked by recent progress in LLMs but must not limit our imaginations and aspirations by only what LLMs make plausible.
We must also consider the broad sociotechnical implications of deploying these emerging technologies in the context of information access~\citep{mitra2024sociotechnical}, and their systemic consequences and risks.

\paragraph{Our relationships with adjacent communities are changing and so is how we do research.}
Not long ago, many in the IR community would tout IR, and specifically web search, as a rare success story of real-world application of ML, AI, and NLP technologies.
A sea change in these adjacent communities in the last decade have shifted this balance.
Now, many see IR as just another NLP task, sometimes included in NLP benchmarks (\eg, HELM~\citep{liang2022holistic}) for evaluating ML and AI models.
With retrieval-augmented LLMs, IR is auditioning for a new role, as a tool for AI models, curiously inverting the relationship between these technologies, where AI was one of many in the IR toolbox.

These communities are also undergoing significant changes in research culture, often influencing each other on the way.
One particular trend in NLP, and the broader ML and AI communities, that has influenced IR, is \emph{leaderboard-driven research}.
Several NLP leaderboards~\citep{bajaj2016ms, goyal2017making, joshi2017triviaqa, rajpurkar2018know, wang2018glue, talmor2018commonsenseqa, wang2019superglue, kwiatkowski2019natural, yang2018hotpotqa, liang2020xglue} have been instrumental in encouraging progress on specific tasks.
IR has a long history of focus on shared tasks and benchmarks, notably TREC~\citep{voorhees2005trec} that has been a venue for developing new tasks and benchmarks, as well as building research communities with shared interests around them.
What differentiates IR benchmarking in venues such as TREC from NLP leaderboards is that the former is framed not as a \emph{competition}, but as a \emph{coopetition}.
In a competition, the goal of the participant is to outperform others, while in a coopetition the participants share a collective goal to develop a better understanding of both the task and the models in the spirit of scientific enquiry.
In the words of IR researcher Ian Soboroff:\footnote{\url{https://twitter.com/ian_soboroff/status/1426901262369439751}} ``The datasets were not built to be solved. They were built as tools to understand the problem and the systems we build to `solve' them.''; or as \citet{voorhees2021coopetition} put it: ``Coopetition is defined as competitors cooperating for the common good... While competition can give one a bigger piece of the pie, cooperation makes the whole pie bigger.''
By emphasizing the goals of community development and understanding of the tasks and the models, these evaluation effort try to promote scientific enquiry over sportive competition.  
Even the MS MARCO ranking task that initially started as a competition later reframed itself as a coopetitive evaluation effort~\citep{craswell2021ms, lin2022fostering}.

As a community we should cautiously embrace insights and trends from our neigboring fields.
However, we should not let IR be minimized to just an evaluation task which undermines the critical responsibility that IR researchers owe to the broader society.
Similarly, while leaderboards and competitions may be effective in creating excitement and increasing participation in certain tasks, we must be mindful of the implications of potentially a large section of the IR community being driven predominantly by these practices.
When the goal is to win, then scientific inquiry takes a back seat, and the ones with the most compute and data resources take the metaphorical steering wheel.
It risks, what \citet{gausen2023framework} call, albeit in a different context, shifting from \emph{praxis}---\ie, ``reflection and action directed at the structures to be transformed''~\citep{freire2020pedagogy}---to \emph{proxies}, \ie, optimizing towards proxy quantitative measures of outcomes.
Actions in this context may refer to any research activity, including but not limited to: formalization, design, experimentation, publishing, artefact creation, open sourcing, and community building; and examples of proxies include state-of-the-art (SOTA) performance on benchmarks and leaderboard rankings that do not translate to better scientific understanding or positive impact on people.\footnote{This is a case of Goodhart's law~\citep{chrystal01goodhart,goodhart75monetary,hoskin96awful,thomas22reliance} whereby improvements on benchmarks and corresponding metrics do not translate to progress on the problem the benchmark was created for, as has been argued for example by \citet{hsia2023goodhart}.}

Yet another relationship that we must critically examine is the one between industry and academia.
\citet{whittaker2021steep} point out that the concentration of data and compute resources, two key ingredients in recent advances in AI, in the hands of few large tech corporations is giving these same institutions tremendous power to shape academic research agenda.
Big tech also shapes academic research agendas in various other ways, including academic engagements and employments.
In IR, the MS MARCO dataset~\citep{bajaj2016ms} and leaderboard~\citep{craswell2021ms, lin2021significant, lin2022fostering}, and the TREC Deep Learning track~\citep{trec2020overview}, that has been broadly adopted for benchmarking deep ranking models were produced and is currently maintained by industry researchers.
Indeed, the organizers behind these efforts themselves recognized~\citep{craswell2021ms, lin2022fostering} the critical responsibility that comes with defining critical research tasks for the community---effectively playing ``the Pied Piper guiding a significant section of the community down specific lanes of research''---and recommend all benchmark developers to engage in open and inclusive discussions with the rest of the community to critically examine their impact.
While academia-industry collaboration is critically important for the field to ground our research in real large-scale systems and see our research outputs materialize into real impact on system users, we must also resist the homogenization of our research agendas towards a singular world view put forth by Big Tech capitalism. 

\paragraph{The world is changing and so is our relationship to that world.}
Our world at large is experiencing a confluence of many simultaneous, and mutually reinforcing, forces that are increasingly pushing us towards precarity, including but not limited to: increasing global wealth and income inequality~\citep{chancel2022world}, rising global conflicts~\citep{taylor2023historic, un2023highest}, pandemics~\citep{scientist2021covid, taylor2022covid, centers2022cdc}, and impending climate catastrophes~\citep{parmesan2022climate, ipcc2013physical, poynting2024confirmed}.
At a moment when the world needs global solidarity built on trust and consensus, and informed citizenry with robust access to reliable information, online disinformation and misinformation are undermining both~\citep{turrentine2022climate, treen2020how, kata2010postmodern, allcott2017social, doubek2017disinformation, beaumont2020malicious, zadrozny2024disinformation, swenson2023social}.
While these complex global challenges require sophisticated and multifaceted response that spans across the political, legal, economic, and technological realms, one thing is for certain that information access research has a role to play.
\emph{So, will we answer the call?}
\section{Conclusion}
\label{sec:conclusion}

\paragraph{Despite good intentions.}
We must be vigilant and reflexively critique our impact, whether under the model of existing fairness and ethics research in IR or under the proposed shift.
The call for pluralistic sociotechnical imaginaries must not in this context be confused with uncritical acceptance of all possible futures as equally valid or desirable.
Instead, this is a call for critical examination of our community's existing normative values and future aspirations.
This work not only involves explicating our sociotechnical imaginaries but also engaging critically with the history of technology---\eg,~\citep{merchant2023blood}---and challenging harmful silicon valley ideologies that are counter to the goals of universal emanicipation and justice---\eg,~\citep{gebru2023eugenics}.
Above all, we should be wary of any promises of the future that further concentrates wealth and power, or advances any notion of altruism in place of structural change~\citep{giridharadas2019winners}.

\paragraph{Desired outcomes.}
Having emphasized the importance of theories of change and ensuring that our research has the desired societal impact and not merely constitute an intellectual exercise, it is only fair that we explicate our own desired outcome of this particular work.
We authored this paper because we sincerely believe that information access has a critical role to play in determining our collective futures; and that real change can not be realized by fairness and ethics research happening in silos but only when combined with raising social consciousness, organizing, and movement building.
We would consider it a failure if this paper is only cited in future IR papers as a passing remark on social responsibility of IR research.
Instead, we hope this work sparks many passionate conversations and debates within the community, and radicalizes us to work on issues of social import in collaboration with other disciplines and civil society.
But above all, we hope this paper serves as a clarion call to all IR researchers to reflect on why we do what we do.
Personally, we hope that the community continues to build technology not just because we love technology itself, but as an act of radical love for all peoples and the worlds we share.
So, we conclude with one final quote for our readers.

\begin{quote}
  ``Another world is not only possible, she is on her way. On a quiet day, I can hear her breathing.''
  \begin{flushright}
  -- Arundhati Roy\\
  \emph{War talk}~\citep{roy2003war}
  \end{flushright}
\end{quote}

\subsection*{Positionality statement}
The author of this paper works at a large technology company in the global north.
However, the perspectives presented in this work is intended to challenge Big Tech and global north's view of technology and our collective futures.
\acks{The author gratefully acknowledges feedback from Asia Biega, Michael D. Ekstrand, and Ida Larsen-Ledet on the various drafts of this paper. No external funding was received in support of this work.}

\vskip 0.25in


\end{document}